\documentclass[aps,twocolumn,showpacs,nofootinbib,groupedaddress]{revtex4}
\usepackage{graphicx}
\usepackage{amsmath}
\usepackage{amsfonts}
\usepackage{amssymb}
\usepackage{hyperref}

\begin{document}

\title{Discrete Wigner functions and quantum computational speedup}

\author{Ernesto F. Galv\~{a}o}
\affiliation{Perimeter Institute for Theoretical Physics \\31
Caroline Street North, Waterloo, Ontario, N2L 2Y5, Canada}

\begin{abstract}
Gibbons \textit{et al.} [\textit{Phys. Rev. A} \textbf{70}, 062101
(2004)] have recently defined a class of discrete Wigner functions
$W$ to represent quantum states in a finite Hilbert space
dimension $d$. I characterize the set $C_d$ of states having
non-negative $W$ simultaneously in all definitions of $W$ in this
class. For $d\le5$ I show $C_d$ is the convex hull of stabilizer
states. This supports the conjecture that negativity of $W$ is
necessary for exponential speedup in pure-state quantum
computation.
\end{abstract}
\date{\today}
 \maketitle

\section{Introduction}

Continuous-variable quantum systems can be represented in phase
space using various quasi-probability distributions, notably the
Wigner function $W(q,p)$ \cite{Wigner32,HillaryOSW84}. This
real-valued function plays some of the roles of the classical
Liouville density, for example allowing us to calculate some
system properties through phase-space integrals weighted by
$W(q,p)$. Despite these similarities, $W(q_0,p_0)$ cannot be
interpreted as the probability of simultaneously measuring
observables $\hat{p}$ and $\hat{q}$ with eigenvalues $p_0$ and
$q_0$: such dispersion-free values for non-commuting observables
are not allowed in quantum mechanics. In fact, $W(q,p)$ can even
be \textit{negative} in some phase-space regions, something that
obviously could not happen were $W(q,p)$ a true probability
distribution (hence the term \textit{quasi-}probability).

The connection between negativity of $W(q,p)$ and non-classicality
has not been completely fleshed out, partly due to different
subjective views on what qualifies as `non-classical behavior' (see \cite{Hudson74,BuzekV-DK92}).
Negativity of $W(q,p)$ has been linked to non-locality
\cite{Bell86}, but the relation is not straightforward. For
example, the original Einstein-Podolsky-Rosen state can display
non-locality despite having positive $W(q,p)$ in all phase space
\cite{Cohen97,BanaszekW98,BanaszekW98b}.

Buot \cite{Buot74} and Hannay and Berry \cite{HannayB80} seem to
have been among the first to propose analogues of the Wigner
function for finite-dimensional Hilbert spaces. Their finds were
rediscovered later by Cohen and Scully \cite{CohenS86} and Feynman
\cite{Feynman87}, who defined a discrete Wigner function $W$ for
the case of a single qubit. This work was developed by Wootters
\cite{Wootters87} and Galetti and de Toledo Piza
\cite{GalettiP88}, who introduced a Wigner function  for
prime-dimensional Hilbert spaces. There followed other definitions
valid for dimension $d$ which is odd \cite{CohendetCSS-C88}, even
\cite{Leonhardt95}, power-of-prime \cite{Wootters04, GibbonsHW04},
or arbitrary \cite{RivasOdA99,BianucciMPS02,MiquelPS02}. These
have been recently used to visualize and get insights on
teleportation \cite{KoniorczykBJ01,Paz02}, quantum algorithms
\cite{BianucciMPS02,MiquelPS02}, and decoherence \cite{LopezP03}.
A recent review of phase-space methods for finite-dimensional
systems is given in \cite{Vourdas04}.

In this paper I investigate the relation between negativity of
discrete Wigner functions and quantum computational speedup. I
will focus on the class of Wigner functions defined by Wootters
\cite{Wootters04} and Gibbons \textit{et al.} \cite{GibbonsHW04}
for power-of-prime dimensions. Wigner functions in this class are
defined by associating lines in a discrete phase space to
projectors belonging to a fixed set of mutually unbiased bases. We
will see that the set of states displaying negativity of $W$
depends on a number of arbitrary choices required to pick a
particular definition of $W$ from the class. I eliminate this
arbitrariness by characterizing the set $C_d$ of states having
non-negative $W$ simultaneously in \textit{all} definitions in the
class, for $d$-dimensional Hilbert space.

For a single qubit, I show that the set $C_2$ consists of states
which fail to provide quantum computational advantage in a model
recently proposed by Bravyi and Kitaev \cite{BravyiK04}. For
dimensions $2\le d \le 5$, I show that the set $C_d$ is the convex
hull of a set of \textit{stabilizer} states, i.e. simultaneous
eigenstates of generalized Pauli operators
\cite{Gottesman97,Gottesman99b}. This is interesting, as quantum
computation which is restricted to stabilizer states and gates
from the Clifford group can be simulated efficiently on a
classical computer \cite{Gottesman99}. If the result holds for
arbitrary power-of-prime dimensions (as I conjecture), then pure
states in $C_d$ would be `classical' in the sense of having an
efficient description using the stabilizer formalism. This
suggests that states with negative $W$ may be \textit{necessary}
for exponential quantum computational speedup with pure states.

\section{Discrete Wigner functions}

In this section we review the discrete Wigner functions introduced
by Wootters \cite{Wootters87} and Galetti and de Toledo Piza
\cite{GalettiP88} for prime dimensions, and elaborated on recently
by Wootters \cite{Wootters04} and Gibbons, Hoffman and Wootters
\cite{GibbonsHW04} for prime-power dimensions. We start by
defining discrete analogues of phase space and its partitions into
parallel `lines' (i.e., \textit{striations}). Then we review how
to define a class of discrete Wigner functions $W$ by associating
lines with projectors onto basis vectors from a set of mutually
unbiased bases (MUB's).

\subsection{Phase space and striations}

The discrete analogue of phase space in a $d$-dimensional Hilbert
space is a $d \times d$ real array. Unlike the continuous phase
space, in this discrete setting we do not have geometrical lines
of points. Instead, a `line' is defined as a set of $d$ points in
discrete phase space.

It is clear that our discrete phase space can then be partitioned
in multiple ways into collections of parallel lines (i.e. disjoint
sets of $d$ phase-space points). Following Wootters
\cite{Wootters04}, we call each such partition a
\textit{striation}.

In \cite{GibbonsHW04} a procedure for building $(d+1)$ striations
of a $d\times d$ phase-space array was outlined, resulting in
striations with the following three useful properties:
\begin{description}
\item[i] Given any two points, exactly one line contains both points;\label{pstria1}
\item[ii] given a point $\alpha$ and a line $\lambda$ not containing $\alpha$, there is exactly one line parallel to $\lambda$ that contains $\alpha$;\label{pstria2}
\item[iii] two non-parallel lines intersect at exactly one point. \label{pstria3}
\end{description}
The construction of these $(d+1)$ special striations involves
labeling the discrete phase space with elements of finite fields,
and defining lines using natural properties of the field (for
details, see \cite{GibbonsHW04}). These striations play a central
role in the definition of the discrete Wigner function $W$, as we
will see below.

\subsection{Mutually unbiased bases}

Wootters' definition of discrete Wigner functions makes use of a
special set of $d+1$ bases for a $d$-dimensional Hilbert space.
Consider two different orthonormal bases $B_1$ and $B_2$:
\begin{eqnarray}
B_1=\{
\left|\alpha_{1,1}\right\rangle,\left|\alpha_{1,2}\right\rangle,...,\left|\alpha_{1,d}\right\rangle
\}, \left|\left\langle
\alpha_{1,i}|\alpha_{1,j}\right\rangle\right|^2 =\delta_{i,j}
,\\
B_2=\{
\left|\alpha_{2,1}\right\rangle,\left|\alpha_{2,2}\right\rangle,...,\left|\alpha_{2,d}\right\rangle
\}, \left|\left\langle
\alpha_{2,i}|\alpha_{2,j}\right\rangle\right|^2
=\delta_{i,j}.\end{eqnarray} These two bases $B_1$ and $B_2$ are
said to be \textit{mutually unbiased}, or \textit{mutually
conjugate}, if
\begin{equation}
\left|\left\langle
\alpha_{i,j}|\alpha_{k,l}\right\rangle\right|^2=\frac{1}{d} \text{
if } i\neq k.
\end{equation}

Wootters and Fields showed that one can define $(d+1)$ such
mutually unbiased bases (MUB's) for power-of-prime dimension $d$
\cite{WoottersF89}. Note that this is exactly the number of
striations one can find with properties \textbf{i}-\textbf{iii}
above; we will use this now to define a class of discrete Wigner
functions \textit{W}.

\subsection{Defining a class of discrete Wigner functions}

We now have the ingredients to define a class of discrete Wigner
functions: a set of $(d+1)$ mutually unbiased bases $\{B_1,
B_2,...,B_{d+1}\}$; and a set of $(d+1)$ striations
$\{S_1,S_2,...,S_{d+1}  \}$ of our $d \times d$ phase space into
$d$ parallel lines (i.e. disjoint sets) of $d$ points each. To
define a discrete Wigner function we need to choose two one-to-one
maps:

\begin{itemize}
\item each basis set $B_i$ is associated with one striation $S_i$; and
\item each basis vector $\left|\alpha_{i,j}\right\rangle$ is associated with a line $\lambda_{i,j}$ (the $j$th line of the $i$th striation).
\end{itemize}
With these associations, the Wigner function $W$ is uniquely
defined if we demand that
\begin{equation}
Tr\left(
\left|\alpha_{i,j}\right\rangle\left\langle\alpha_{i,j}\right|
\hat{\rho} \right)=\sum_{\alpha\in
\lambda_{i,j}}W_\alpha,\label{defW}
\end{equation}
i.e. we want the sum of the Wigner function elements corresponding
to each line to equal the probability of projecting onto the basis
vector associated with that line.

Note that there are multiple ways of making these associations. In
general, this will lead to different definitions of $W$ using the
same fixed set of MUB's. The procedure outlined above then leads
not to a single definition of $W$, but to a class of Wigner
functions instead.

\section{Wigner function for a qubit}

Let us illustrate Wootters' definition of a discrete Wigner
function with the simplest case, a qubit. The discrete Wigner
function $W$ is defined on a $2 \times 2$ array:
\begin{equation}
W=
\begin{tabular}{|c|c|}\hline
$W_{1,1}$ & $W_{1,2}$ \\ \hline $W_{2,1}$ & $W_{2,2}$  \\\hline
\end{tabular}
\end{equation}

There are three striations of this phase space with properties
\textbf{i,ii} and \textbf{iii} as required. Below I list them,
using numbers $j=1,2$ to arbitrarily label the two lines
$\lambda_{i,j}$ in each striation $S_i$:
\begin{equation}
S_1:
\begin{tabular}{|c|c|}\hline
1 & 1 \\ \hline 2 & 2  \\\hline
\end{tabular},
S_2:
\begin{tabular}{|c|c|}\hline
1 & 2 \\ \hline 1 & 2  \\\hline
\end{tabular}
,S_3:
\begin{tabular}{|c|c|}\hline
1 & 2 \\ \hline 2 & 1  \\\hline
\end{tabular}.
\end{equation}

Now we need to define a set of 3 mutually unbiased bases for a
qubit. These can be conveniently chosen as the eigenstates of the
three Pauli operators $\hat{\sigma}_x, \hat{\sigma}_y$ and
$\hat{\sigma}_z$. Let us label these basis vectors
$\left|\alpha_{i,j}\right\rangle$, where $i\in\{1,2,3\}$ indexes
the MUB ($i=1,2,3$ respectively for the operators
$\hat{\sigma_x},\hat{\sigma_y},\hat{\sigma_z}$), and $j\in\{1,2\}$
indexes the basis vector in each MUB ($j=1,2$ indicates
eigenstates with eigenvalues respectively equal to $+1,-1$).

We can now define $W$ by imposing condition (\ref{defW}), i.e. we
want the sum of $W_{i,j}$ in each line $\lambda$ to be the
probability $p_{i,j}$ of projecting the state onto the basis
vector $\left|\alpha_{i,j}\right\rangle$:
\begin{equation}
p_{i,j}\equiv Tr\left(
\left|\alpha_{i,j}\right\rangle\left\langle\alpha_{i,j}\right|
\hat{\rho} \right)=\sum_{\alpha \in
\lambda_{i,j}}W_{\alpha}.\label{trformula}
\end{equation}

Using the three striations we defined, these conditions can be
explicitly stated, in terms of the probabilities $p_{i,j}$:
\begin{eqnarray}
W_{1,1}+W_{1,2}&=&p_{1,1},\\
W_{2,1}+W_{2,2}&=&p_{1,2},\\
W_{1,1}+W_{2,1}&=&p_{2,1},\\
W_{1,2}+W_{2,2}&=&p_{2,2},\\
W_{1,1}+W_{2,2}&=&p_{3,1},\\
W_{1,2}+W_{2,1}&=&p_{3,2}.
\end{eqnarray}

The equations above define uniquely the Wigner function $W$ in
terms of the probabilities $p_{i,j}$:
\begin{eqnarray}
W_{1,1}=\frac{1}{2}\left( p_{1,1}+p_{2,1}+p_{3,1}-1\right), \label{w2a}\\
W_{1,2}=\frac{1}{2}\left( p_{1,1}+p_{2,2}+p_{3,2}-1\right),\\
W_{2,1}=\frac{1}{2}\left( p_{1,2}+p_{2,1}+p_{3,2}-1\right),\\
W_{2,2}=\frac{1}{2}\left(
p_{1,2}+p_{2,2}+p_{3,1}-1\right).\label{w2d}
\end{eqnarray}

Here we should keep in mind that not all probabilities $p_{i,j}$
are independent, as the sum over any MUB must add up to $1$ (i.e.
$\forall i, \sum_j {p_{i,j}=1}$).

From the definitions (\ref{w2a})-(\ref{w2d}) we can immediately
read the conditions for non-negativity of the Wigner function:
\begin{eqnarray}
p_{1,1}+p_{2,1}+p_{3,1}\ge1 \label{ip2a}\\
p_{1,1}+p_{2,2}+p_{3,2}\ge1\\
p_{1,2}+p_{2,1}+p_{3,2}\ge1\\
p_{1,2}+p_{2,2}+p_{3,1}\ge1\label{ip2d}
\end{eqnarray}
Because of the conditions $\sum_j {p_{i,j}=1}$, there are only
three free variables in the inequalities above. We can thus pick
one $p_{i,j}$ for each MUB $i$ (say, $p_{i,1}$), and represent
states by points in this three-dimensional probability space. This
representation is equivalent to the more familiar Bloch sphere,
with the advantage of generalizing easily to power-of-prime
dimension $d$. In that case (to be discussed in section
\ref{sechighd}), there are a total of $(d^2-1)$ independent
probabilities $p_{i,j}$ describing any mixed quantum state. Note
that the probabilities $p_{i,j}$ combine linearly for convex
combinations (probabilistic mixtures) of quantum states, a
consequence of the linearity of the trace in eq.
(\ref{trformula}).

The inequalities (\ref{ip2a})-(\ref{ip2d}) for non-negativity of
$W$ are satisfied by state $\vec{p}=(p_{1,1},p_{2,1},p_{3,1})$ if
and only if $\vec{p}$  lies inside the tetrahedron
\begin{equation}
T_1=\text{convex hull of }\{(0,0,1),(0,1,0), (1,0,0),(1,1,1)\}
\end{equation}
(see figure \ref{fig pqubitpoly}).

By contrast, there exist quantum states whose $W$ is negative in
some phase-space points, i.e. lying outside of tetrahedron $T_1$.
This is illustrated in figure \ref{fig pqubitpoly}, where the set
of points corresponding to quantum states is the ball of radius
$r=1/2$ and with center $(\frac{1}{2},\frac{1}{2},\frac{1}{2})$.

\begin{figure}
\begin{center}
{\includegraphics[scale=0.45]{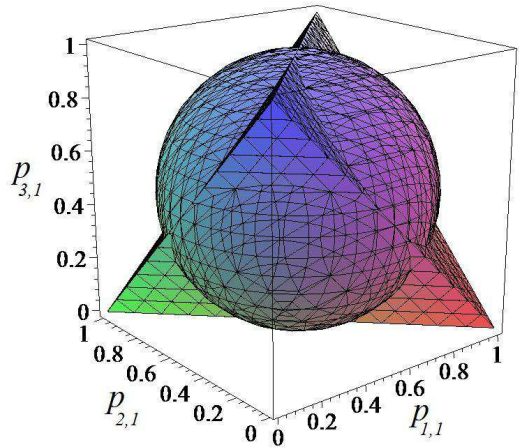}} \caption[One-qubit
non-negative Wigner function tetrahedron.]{\label{fig
pqubitpoly}Tetrahedron in $\vec{p}=(p_{11},p_{21},p_{31})$ space
representing states with non-negative Wigner function $W$, for the
definition of $W$ given by eqs. (\ref{w2a})-(\ref{w2d}). The ball
represents one-qubit quantum states.}
\end{center}
\end{figure}

Let me draw attention to two features of the set of states with
non-negative $W$. First, there exist points $\vec{p}\in T_1$ which
do not correspond to any one-qubit quantum state (e.g.
$\vec{p}=(1,1,1)$). We must also keep in mind that there were
arbitrary choices involved in the particular definition of $W$
that we picked. We have chosen arbitrary one-to-one maps between
MUB's and striations, and also between lines in a striation and
basis vectors. In the next section we take these two points into
account, defining a set which consists solely of quantum states,
and for which the Wigner function is non-negative for \textit{all}
definitions of $W$.

\section{One-qubit states with non-negative Wigner functions}

In this section I use negativity of $W$ to  define the set $C_d$
of states in $d$-dimensional Hilbert space having non-negative $W$
for all definitions of $W$ based on a fixed set of MUB's. I will
then argue that single-qubit states in $C_2$ behave `classically'
in a concrete computational sense.

Having fixed a set of 3 MUB's for a single qubit, Wigner functions
$W$ can be defined in a number of ways, corresponding to the
$3!(2!)^3=48$ different associations between lines and basis
vectors, and between striations and MUB's. For each of these $48$
Wigner function definitions we can do as above, and find the set
of points in $\vec{p}$-space for which $W$ is non-negative. A
simple calculation shows that depending on the definition, this
set is either the original tetrahedron $T_1$ or the tetrahedron
whose vertices are the other four $\vec{p}$-cube vertices:
$T_2=\text{convex hull of }\{(1,1,0),(1,0,1),(0,1,1),(0,0,0)\}$.

Now I define the set $C_d$ of states in $d$-dimensional Hilbert
space which I will argue behave `classically' in a computational
sense:

\begin{description}
\item[Definition:] The set $C_d$ is defined as the states in a $d$-dimensional Hilbert space whose Wigner function $W$ is non-negative in all phase space points \textit{and} for all definitions of $W$ using a fixed set of mutually unbiased bases.
\end{description}
In the remainder of this article I will characterize the set $C_d$
for some small dimensions $d$, and discuss the limitations of
doing quantum computation solely with states in $C_d$.

\begin{figure}
\begin{center}
{\includegraphics[scale=0.40]{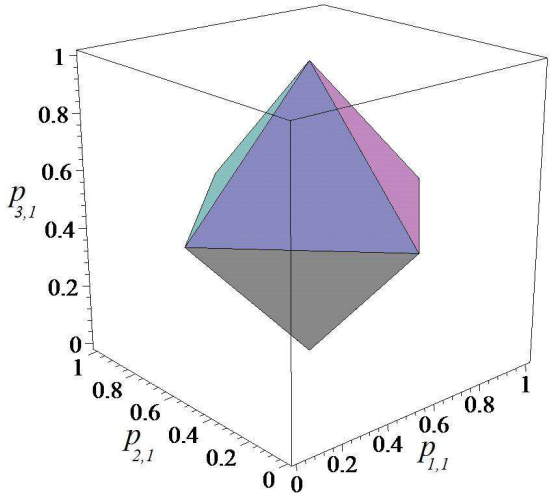}} \caption[Octahedron of
states whose Wigner function is non-negative for all choices of
$W$]{\label{fig octa}Octahedron in
$\vec{p}=(p_{11},p_{21},p_{31})$ space representing states with
non-negative Wigner functions $W$ for all definitions of $W$ using
a fixed set of mutually unbiased bases.}
\end{center}
\end{figure}

For a qubit, the set $C_2$ is given in $\vec{p}$-space as the
intersection of the two tetrahedra $T_1$ and $T_2$ presented
above. This corresponds to an octahedron inscribed inside the ball
of quantum states:
\begin{eqnarray}
 C_2=\text{convex hull of }\{(1,\frac{1}{2},\frac{1}{2}),(0,\frac{1}{2},\frac{1}{2}),(\frac{1}{2},1,\frac{1}{2}),\nonumber \\(\frac{1}{2},0,\frac{1}{2}),(\frac{1}{2},\frac{1}{2},1),(\frac{1}{2},\frac{1}{2},0)\}
\end{eqnarray}
(see figure \ref{fig octa}). We see that set $C_2$ can be
characterized in a simpler way as the set of states which are
convex combinations of the six basis vectors of the three chosen
MUB's:
\begin{equation}
\hat{\rho} \in {C_2} \Leftrightarrow \hat{\rho} =\sum_{i,j}q_{i,j}
\left|\alpha_{i,j}\right\rangle\left\langle\alpha_{i,j}\right|,
\hspace{4 mm }\sum_{i,j}q_{i,j}=1.
\end{equation}

There are at least two motivations for considering the set $C_2$,
as opposed to states with non-negative $W$ in a single definition
(the sets $T_1$ or $T_2$). The first one is the realization that a
priori there is no preferred definition of $W$ from the full class
of definitions, and any concept of `classical states' based on
non-negativity of $W$ should be definition-independent. The second
motivation only became apparent after we calculated $C_2$: unlike
the sets $T_1$ and $T_2$, all states in $C_2$ are physical states,
i.e. obtainable from measurements on a single qubit.

In the next section we will review other properties of $C_2$ from
a computational perspective.

\subsection{States in $C_2$ and quantum computing}

In the last section I defined the set $C_2$ of single-qubit
quantum states which have non-negative $W$ in all definitions of
$W$ using a fixed set of MUB's. Let us now review an argument for
the `classicality' of the set $C_2$, from a computational point of
view.

Recently Bravyi and Kitaev \cite{BravyiK04} proposed the following
model of computation. Imagine that for some reason there are a few
quantum computational operations which we can perform perfectly --
let us denote these by $O_{ideal}$. In addition to those, some
operations in set $O_{faulty}$ can only be performed imperfectly.
Now let us consider a particular choice for $O_{ideal}$:
\begin{itemize}
\item Prepare a qubit in state $\left|0\right\rangle$ (i.e., an eigenstate of the $\hat{\sigma}_z$ operator) ;
\item Apply unitary operators from the Clifford group (such as the Hadamard and CNOT gates);
\item Measure an eigenvalue of a Pauli operator ($\hat{\sigma}_x, \hat{\sigma}_y$ or $\hat{\sigma}_z$) on any qubit.
\end{itemize}
The Gottesman-Knill theorem \cite{Gottesman97} states that the
operations in $O_{ideal}$ above can only create a restricted set
of states known as \textit{stabilizer states}, i.e. simultaneous
eigenstates of the Pauli group of operators. Moreover, such
operations do not allow for universal quantum computation, and can
be efficiently simulated on a classical computer.

In addition to these perfect operations, Bravyi and Kitaev
proposed a set $O_{faulty}$ with a single extra imperfect
operation:
\begin{itemize}
\item Prepare an auxiliary qubit in a mixed state $\hat{\rho}$.
\end{itemize}

In this model of computation, it is easy to see that auxiliary
qubits in states $\hat{\rho} \in C_2$ cannot be used to perform
universal quantum computation \cite{BravyiK04}. States in $C_2$
are linear convex combinations of the six eigenstates of single
Pauli operators. As such, they can be prepared efficiently from
operations in $O_{ideal}$, namely, Clifford gates to obtain any
Pauli eigenstate deterministically from the initial
$\hat{\sigma}_z$ eigenstate, together with classical coin tosses
to prepare them with the appropriate weights given by the desired
convex combination. These are operations that can be efficiently
simulated on a classical probabilistic computer. This justifies my
`classicality' claim for states in $C_2$.

Interestingly, in \cite{BravyiK04} it was shown that some
`non-classical' states $\hat{\rho} \not\in C_2$ can be used to
attain universal quantum computation in this model. The basic idea
is to use auxiliary pure states outside of $C_2$ to implement
gates outside of the Clifford group, using operations in
$O_{ideal}$ only. A single generic non-Clifford gate together with
the set of Clifford gates allows for universal quantum
computation. This procedure works also for a large class of mixed
states $\hat{\rho}\not\in C_2$, through a distillation of pure
non-stabilizer pure states from $\hat{\rho}^{\otimes N}$ using
Clifford operations only \cite{BravyiK04}.

This enables us to identify non-stabilizer states as a resource
which can be tapped to implement non-Clifford gates and achieve
universal quantum computation. This idea was first suggested by
Shor \cite{Shor96}, and has since been elaborated on by other
authors \cite{Dennis99, GottesmanC99, ZhouLC00, JorrandP04,
ChildsLN04}.

\section{Higher dimensions\label{sechighd}}

We can follow Gibbons \textit{et al.} and define a discrete Wigner
function $W$ whenever the Hilbert space dimension $d$ is a power
of prime. In this section I use the definition of $C_d$ to
characterize this set for states in small Hilbert space dimensions
$d$.

In $d$ dimensions, the probability-space point $\vec{p}$
describing each state has $(d^2-1)$ components. Requiring
non-negativity of $W$ for all definitions will correspond to a set
of inequalities for $\vec{p}$, each delimiting a half-space where
$W$ is non-negative at a particular phase-space point, and for a
particular definition of Wigner function. States in the set $C_d$
are those which satisfy all these inequalities, constituting a
convex polytope.

Any convex polytope admits two descriptions, one in terms of the
half-space inequalities (H-description), and one in terms of its
vertices (V-description). The set $C_d$  is, by definition, an
H-polytope. For one qubit we have found the equivalent
V-description: there are $6$ vertices corresponding to the $6$
basis vectors of the three MUB's used to define $W$.

Let us now see how to find the equivalent V-description for the
H-polytope $C_d$ defined above. The starting point is the general
expression for the $d$-dimensional Wigner function at phase-space
point $\alpha$ (see \cite{GibbonsHW04}):

\begin{equation}
W_{\alpha}=\frac{1}{d}\left[ \sum_{\lambda_{i,j} \ni
\alpha}p_{i,j}-1  \right],\label{eqwgen}
\end{equation}
where the sum is over the probabilities associated with projectors
corresponding to all lines $\lambda_{i,j}$ containing phase-space
point $\alpha$. By construction of the striations, each point
belongs to exactly one line from each striation. Thus for each
point $\alpha$,  the sum above has $(d+1)$ terms $p_{i,j}$, one
from each MUB.

We want to find which conditions on $p_{i,j}$ correspond to the
demand of non-negativity of $W_{\alpha}$ for all points $\alpha$
\textit{and} for all definitions of $W$ using a fixed MUB set. For
a fixed phase-space point $\alpha$, changing the definition of $W$
will correspond to picking a different set of $(d+1)$
probabilities $p_{i,j}$ in eq. (\ref{eqwgen}), one from each MUB
$i$. There are only $d^{(d+1)}$ ways of doing so, and this is the
total number of expressions for $W_{\alpha}$ which we would like
to take only non-negative values. These $d^{(d+1)}$ inequalities
of the form $W_{\alpha}\ge0$ constitute our H-description of the
polytope $C_d$.

From this H-description it is possible to find the equivalent
V-description using a convex hull program based on the QuickHull
algorithm \cite{BarberDH96}. While this is considered to be a
computationally hard problem in general, it was possible to do the
calculation for $d\le5$. The results are similar to the one-qubit
case. For $d=3,4$ and $5$ I found that the V-description of $C_d$
has $d(d+1)$ vertices, each corresponding to one of the MUB basis
vectors used to define $W$. For $d=5$, for example, the H-polytope
$C_5$ in $5^2-1=24$-dimensional $\vec{p}$-space is delimited by
$5^6=15625$ half-space inequalities corresponding to
non-negativity of $W$ in all definitions. The V-description of
this polytope consists of exactly $5 \times 6=30$ vertices, each
corresponding to one of the basis vectors of the six MUB's.

It is not hard to see that the MUB basis vectors have non-negative
$W$ in every single definition of $W$ using those basis vectors.
For $d\le5$, the computation described above shows that these are
the only pure states for which this is true for all definitions,
and moreover that the mixed states with the same property are
exactly their convex combinations. In the next section we will
discuss the implications for quantum computing.

\subsection{The set $C_d$ and quantum computing}

Given the results above for $d=2,3,4$ and $5$, it is natural to
make a conjecture:
\begin{description}
\item[Conjecture:] For any power-of-prime Hilbert space dimension $d$, the polytope $C_d$ is equivalent to the V-polytope whose vertices are the basis vectors of the MUB's used to define $W$.
\end{description}
It is easy to show that the latter V-polytope is contained in
$C_d$; the converse seems to be harder to prove for general
power-of-prime $d$.

To understand the relevance of this conjecture for quantum
computation, we need to review some known facts about mutually
unbiased bases. Various authors have come up with constructions of
MUB's
\cite{Ivanovic81,WoottersF89,BandyopadhyayBRV02,LawrenceBZ02,PittengerR03,KlappeneckerR04}.
For $d=2^N$, i.e. for the case of $N$ qubits, the corresponding
basis vectors can always be written as simultaneous eigenstates of
tensor products of single-qubit Pauli operators
\cite{LawrenceBZ02}. More generally, for any power-of-prime
dimension $d$ the MUB basis vectors can be chosen to be
eigenstates of generalized Pauli operators
\cite{PittengerR00,BandyopadhyayBRV02}. In other words, the basis
vectors of MUB's can always be chosen to be a set of stabilizer
states.

The conjecture would then mean that for any $d$, the set $C_d$
would be the convex hull of the particular subset of stabilizer
states used as basis vectors of the MUB's. Pure-state quantum
computation which is restricted to states in $C_d$ would then only
ever reach the MUB basis vectors, all of which admit an efficient
classical description using the stabilizer formalism.

As I have already remarked, it is well-known that quantum
computation that is restricted to pure stabilizer states and gates
from the Clifford group is not universal, and can be efficiently
simulated on a classical computer \cite{Gottesman97}. If the
conjecture holds, then a wide class of unitary dynamics restricted
to pure states inside $C_d$ would be easy to simulate on a
classical computer -- for example, any gate from the Clifford
group. This suggests that non-trivial pure-state quantum
computation may require states outside of $C_d$, i.e. having
negative Wigner function $W$ in at least some of the definitions
of $W$ in the class proposed by Gibbons, Hoffman and Wootters
\cite{GibbonsHW04}.

\section{Conclusion}

In this paper I related the negativity of the discrete Wigner
function $W$ with quantum computational speedup. I characterized
the set of states with non-negative $W$ using the class of Wigner
functions $W$ proposed by Wootters and collaborators in
\cite{Wootters04,GibbonsHW04}. Wigner functions in this class are
defined using projectors on a fixed set of mutually unbiased bases
(MUB's). The set of states with non-negative $W$ depends on which
definition of $W$ from this class we choose.

I defined the set $C_d$ of $d$-dimensional states having
non-negative $W$ simultaneously in \textit{all} definitions of $W$
in the class. States in $C_2$ are classical in the sense of
failing to provide quantum computational advantage in a recent
model proposed by Bravyi and Kitaev \cite{BravyiK04}. For
dimensions $2 \le d \le 5$ I showed that $C_d$ is the convex hull
of the stabilizer states used as basis vectors in the MUB set.

These results for small dimensions $d$ support the conjecture that
for \textit{any} power-of-prime dimension $d$, the set $C_d$ is
the convex hull of a set of stabilizer states. This would mean
that pure states in $C_d$ are `classical' in the sense of
admitting an efficient classical description using the stabilizer
formalism. This suggests that states with negative $W$ may be
\textit{necessary} for exponential pure-state quantum
computational speedup.

\textbf{Acknowledgements.} I would like to thank Angelo Carollo,
David Poulin, Rob Spekkens, Daniel Gottesman and Lucien Hardy for
helpful discussions. This work was supported in part by Canada's
NSERC.


\end{document}